\documentclass[a4paper,11pt]{article}
\usepackage{pos}
\usepackage{ulem}

\title{Hyperons in neutron star mergers}

\author*[a]{Hristijan Kochankovski}
\author[a]{Angels Ramos}
\author[b,c,d]{Laura Tolos}
\author[e,f]{Sebastian Blacker}
\author[f,g]{Andreas Bauswein}

\affiliation[a]{Departament de F\'{\i}sica Qu\`antica i Astrof\'{\i}sica and Institut de Ci\`encies del Cosmos, Universitat de Barcelona, Mart\'i i Franqu\`es 1, 08028, Barcelona, Spain}

\affiliation[b]{Institute of Space Sciences (ICE, CSIC), Campus UAB, Carrer de Can Magrans, 08193 Barcelona, Spain}

\affiliation[c]{Institut d'Estudis Espacials de Catalunya (IEEC), 08860 Castelldefels (Barcelona), Spain}

\affiliation[d]{Frankfurt Institute for Advanced Studies, Ruth-Moufang-Str. 1, 60438 Frankfurt am Main, Germany}

\affiliation[e]{Institut f\"ur Kernphysik, Technische Universit\"at Darmstadt, 64289 Darmstadt, Germany}
\affiliation[f]{GSI Helmholtzzentrum f\"ur Schwerionenforschung, Planckstra{\ss}e 1, 64291 Darmstadt, Germany}
\affiliation[g]{Helmholtz Research Academy Hesse for FAIR (HFHF), Campus Darmstadt, 64291 Darmstadt, Germany}

\emailAdd{hriskoch@fqa.ub.edu}

\abstract{
We discuss the effects induced by the potential presence of hyperons in hot and ultra-dense matter within the context of neutron star mergers. Specifically, we address their effect on the dominant post-merger frequency of the gravitational waves. By performing a simulation campaign with a large sample of hyperonic and nucleonic equations of state, we explicitly show that the unique thermal behavior of hyperonic equations of state results in a systematic shift of the dominant frequency with respect to the nucleonic reference level. The predicted shift has values of up to 150 Hz, and it could be detected with the newest generations of gravitational wave detectors. Thus this approach opens a new path for signaling the presence of hyperons in neutron star remnant matter.

}

\FullConference{10th International Conference on Quarks and Nuclear Physics (QNP2024)\\
8-12 July, 2024\\
Barcelona, Spain\\}

\begin{document}
\maketitle

\section{Introduction}

In the past decades, one of the most intriguing questions that has fascinated nuclear astrophysicists has been 'Are exotic baryons present in ultra-dense matter of neutron stars?' Since the central density of these compact objects can exceed saturation density several times, models describing matter under these conditions have significant uncertainties. Consequently, the composition of the inner core of neutron stars is still unknown, and some models predict that the presence of exotic baryons, such as hyperons and delta baryons, is energetically favorable.

The effects of hyperons on the cold equation of state (EoS) have been extensively studied in the literature \cite{Tolos:2020aln,Burgio:2021vgk,Sedrakian2022,Logoteta:2021iuy,Chatterjee:2015pua}. In recent years, special attention has been devoted to studying their effects also at finite temperature \cite{Oertel:2016xsn,Schaffner-Bielich:2020psc,Motta:2022nlj}. The thermal pressure plays an important role in systems like core-collapse supernovae or neutron star mergers, where the temperature can rise up to tens of MeV and matter moves out of beta equilibrium. Thermal excitations further enhance the possible production of hyperons. It was shown that the decrease of thermal pressure in matter at densities where hyperons become significantly abundant is a clear feature that distinguishes hyperonic matter from nucleonic one \cite{Raduta:2022elz,Kochankovski2022,Kochankovski2024}.   

 With the development of new gravitational wave detectors \cite{Punturo:2010zz, LIGOScientific:2016wof}, in the following decades, we will be able to detect gravitational waves from the post merging phase of binary neutron star mergers. These observables are crucial because they originate from a phase where the matter is hot, making them sensitive to the finite temperature EoS. As hyperons leave traces on the EoS, the observables obtained from these events could carry imprints of hyperons that can help us answer the long lasting question  of their presence in neutron stars.

In this contribution we present a new possible way to get a handle on the composition of dense matter by measuring the dominant post merger gravitational wave frequency from a merger event \cite{Blacker:2023opp}. The analysis was conducted by considering a wide range of hyperonic and nucleonic EoSs and performing an extensive simulation campaign. A distinctive increase in the dominant post-merger frequency is observed in the presence of hyperons, a feature we attribute to the behavior of the thermal pressure and thermal index of hyperonic EoSs. The simulation setup and the results are explained in the following sections.   

\section{Results}


As mentioned, the role of hyperons in cold, evolved neutron stars has been extensively studied, with attempts to extract information on the composition of matter based solely on cold star observables like the mass-radius relation. However, the cold EoSs of hyperonic and nucleonic matter can be very similar such that observables from evolved and isolated neutron stars may not be able to provide a definitive answer to the question of their composition. In the context of quark matter, this issue was dubbed ``masquerade problem'' (see \cite{Alford_2005}) but it obviously applies similarly to hyperonic matter. Considering these challenges, in our approach, we assume that the cold EoS is known with high accuracy, but the composition of the matter still remains unknown. We highlight the fact that the finite temperature behavior of hyperonic and nucleonic matter differ, which we exploit in this work.

We focus on extracting the dominant post merger gravitational wave frequency performing numerous numerical simulations. To ensure our conclusions are systematic, we use a large number of EoSs that consider hyperons as a degree of freedom, but we also use a large sample of nucleonic only EoSs. We refer to both groups as hyperonic and nucleonic sets of EoSs, respectively. 

The basic idea is sketched as follows. First, for each EoS from both sets we perform a numerical simulation of of a 1.4-1.4~$M_\odot$ binary merger, using as input the full temperature-dependent hyperonic and nucleonic EoSs. With this type of simulation, for each EoS, the dominant frequency $f_{peak}$ is obtained. Then, for each EoS from both sets, a reference simulation is made, where instead of using the full temperature-dependent EoS as input, the zero temperature slice in neutrinoless $\beta$-equilibrium of the corresponding EoS is used. This slice is then supplemented with an ideal-gas thermal treatment given by the relation:
\begin{equation}
    P_{th} = \epsilon_{th}(\Gamma_{th} - 1),
\end{equation}
where $P_{th}$ and $\epsilon_{th}$ are the thermal pressure and thermal energy per volume (thermal energy density), respectively, and $\Gamma_{th}$ has a constant value independent of temperature and matter composition. To resemble  the thermal behavior of purely nucleonic matter, the value  $\Gamma_{th} = 1.75$ is used \cite{PhysRevD.88.044026, refId0}. With this second type of simulation, the dominant frequency $f_{peak}^{1.75}$ is obtained. Then, the difference between the two frequencies is calculated:
\begin{equation}
    \Delta f = f_{peak} - f_{peak}^{1.75}.
\end{equation}
The quantity $ \Delta f$ quantifies a frequency shift and serves as a measure of how much the thermal behaviour of a given EoS deviates from an "idealized nucleonic" behaviour. For more details on the EoSs used, the procedure, initial setup, and the numerical treatment, we refer to \cite{Blacker:2023opp} and references therein.
\begin{figure}
    \centering
    \includegraphics[width=0.75\linewidth]{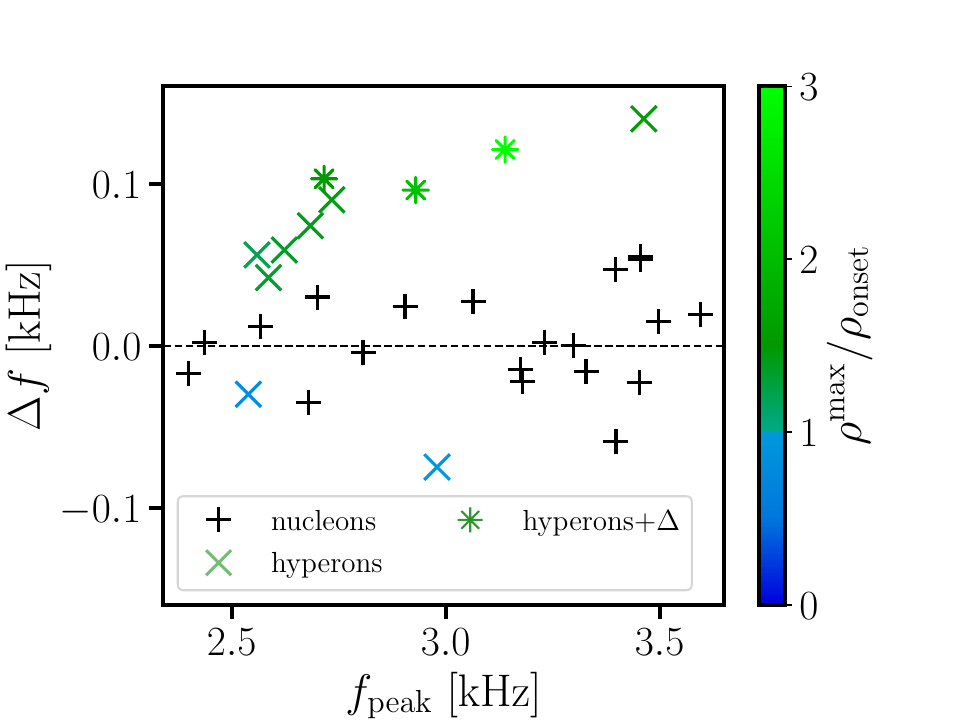}
    \caption{Frequency shift $\Delta f$ as a function of the dominant frequency $f_{peak}$ obtained with temperature dependent EoSs. Models which only consider nucleons are shown by plus signs. Hyperonic models are displayed by crosses. Asterisks denote models which also account for the presence of delta resonances. The color indicates the ratio between the maximum density reached in the remnant and the onset density of hyperons in cold beta equilbirated matter. The figure is taken and adapted from \cite{Blacker:2023opp}.}
    \label{fig:1}
\end{figure}

Fig. \ref{fig:1} shows the frequency shift $\Delta f$ as a function of $f_{peak}$ for all models used in our work. We observe that the hyperonic and nucleonic sets exhibit different behavior. While nucleonic models are scatter around the reference value $\Delta f =0$, hyperonic models show a systematic shift toward positive values. This shift is EoS dependent and can reach values of up to  $\approx 150$ Hz. Furthermore, if the density in the remnant is low (relative to the onset density of hyperons in cold matter for the given EoS) and hyperons are not produced in significant quantities in the remnant, the hyperonic models behave as nucleonic ones. 

\begin{figure}
    \centering
    \includegraphics[width=0.75\linewidth]{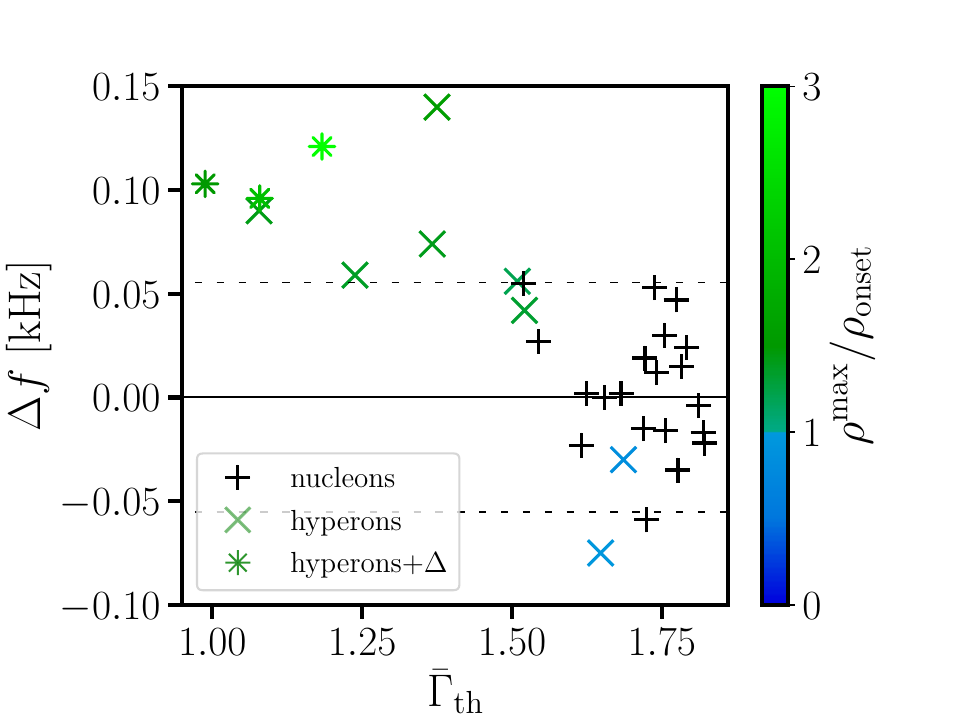}
    \caption{Frequency shift $\Delta f$ as a function of average thermal index of matter $\bar{\Gamma}_{th}$. The labeling is the same as in Fig \ref{fig:1}. The figure is taken and adapted from \cite{Blacker:2023opp}.}
    \label{fig:2}
\end{figure}

To further corroborate the argument that this shift is caused by the distinct thermal behaviour, Fig. \ref{fig:2} shows $\Delta f$ as a function of the mass and time averaged thermal index in the remnant  $\bar{\Gamma}_{th}$ (in a time window of 5 ms staring 2.5 ms after the merging). Please refer to \cite{Blacker:2023opp} for details. 
We can see that there is a correlation between the average thermal index and frequency shift. Hyperonic models cluster in the upper left corner of the graph, corresponding to significant shifts and low average thermal indexes. Conversely, nucleonic models are located in the lower right corner of the graph, corresponding to larger average thermal indexes and lower shifts.

Finally, we discuss some limitations of the study. It is important to note that a frequency shift incompatible with purely nucleonic matter can also be generated by other forms of exotic matter, such as quark matter \cite{Blacker}. Moreover, for a practical implementation of a real event data analysis, accurate gravitational wave measurements and a well-constrained zero-temperature EoS are essential. While in particular the latter assumption appears very optimistic, we refer again to the challenge to discriminate nucleonic and hyperonic models even for the case that well measured stellar parameters are available.

\section{Conclusions}
In this contribution we briefly discuss a novel method of inferring the presence of hyperons in binary neutron star matter using gravitational waves. Through a simulation campaign, we demonstrate that the appearance of thermal hyperons inside of remnant matter induces higher dominant post merger frequencies compared to cases when only ordinary nuclear matter is present. We attribute this behavior to the lower average thermal index of matter when hyperons are present. In future work, we will investigate the influence of hyperons in other quantities relevant for merging events, potentially discovering new signatures of their presence.

\section{Acknowledgments}
We acknowledge funds by the State of Hesse within the Cluster Project ELEMENTS supporting H.K.'s visit during which this project was initialized. 
This research has been supported from the projects CEX2019-000918-M, CEX2020-001058-M (Unidades de Excelencia ``Mar\'{\i}a de Maeztu"), PID2022-139427NB-I00 financed by MCIN/AEI/10.13039/501100011033/FEDER, UE, as well as by the EU STRONG-2020 project, under the program H2020-INFRAIA-2018-1 grant agreement no. 824093, and by PHAROS COST Action CA16214. H.K. acknowledges support from the PRE2020-093558 Doctoral Grant of the spanish MCIN/ AEI/10.13039/501100011033/. 
S.B. and A.B. acknowledge support by Deutsche Forschungsgemeinschaft (DFG, German Research Foundation) through Project-ID 279384907 -- SFB 1245 (subproject B07). A.B. acknowledges support by the European Research Council (ERC) under the European Union’s Horizon 2020 research and innovation program through the ERC Synergy Grant HEAVYMETAL No. 101071865 and support by the State of Hesse within the Cluster Project ELEMENTS.
L.T. also acknowledges support from the Generalitat Valenciana under contract CIPROM/2023/59, from the Generalitat de Catalunya under contract 2021 SGR 171, and from the CRC-TR 211 'Strong-interaction matter under extreme conditions'- project Nr. 315477589 - TRR 211.


\begin{thebibliography}{99}

\bibitem{Tolos:2020aln}Tolos, L. \& Fabbietti, L. \textit{Strangeness in Nuclei and Neutron Stars}.
Prog. Part. Nucl. Phys. \textbf{112}  103770 (2020)
\bibitem{Burgio:2021vgk}Burgio, G., Schulze, H., Vidana, I. \& Wei, J. \textit{Neutron stars and the nuclear equation of state}.  Prog. Part. Nucl. Phys. \textbf{120}  103879 (2021)
\bibitem{Sedrakian2022} Sedrakian, A., Li, J. \& Weber, F. \textit{Hyperonization in Compact Stars}.  Astrophysics In The XXI Century With Compact Stars. World Scientific.  153-199 (2022)
\bibitem{Logoteta:2021iuy}Logoteta, D. \textit{Hyperons in Neutron Stars}.  Universe. \textbf{7}, 408 (2021)
\bibitem{Chatterjee:2015pua}Chatterjee, D. \& Vidaña, I. \textit{Do hyperons exist in the interior of neutron stars?}.  Eur. Phys. J. A. \textbf{52}, 29 (2016)
\bibitem{Oertel:2016xsn}Oertel, M., Gulminelli, F., Providência, C. \& Raduta, A. \textit{Hyperons in neutron stars and supernova cores.}  Eur. Phys. J. A. \textbf{52}, 50 (2016)
\bibitem{Schaffner-Bielich:2020psc} Schaffner-Bielich, J. \textit{Compact Star Physics}. Cambridge University Press, \textbf{8}, (2020)

\bibitem{Motta:2022nlj}Motta, T. \& Thomas, A. \textit{The role of baryon structure in neutron stars}. Mod. Phys. Lett. A. \textbf{37}, 2230001 (2022)

\bibitem{Raduta:2022elz}Raduta, A. \textit{Equations of state for hot neutron stars-II. The role of exotic particle degrees of freedom.}  Eur. Phys. J. A. \textbf{58}, 115 (2022)

\bibitem{Kochankovski2022}Kochankovski, H., Ramos, A. \& Tolos, L. \textit{Equation of state for hot hyperonic neutron star matter.} Mon. Not. Roy. Astron. Soc. \textbf{517}, 507-517 (2022)

\bibitem{Kochankovski2024}Kochankovski, H., Ramos, A. \& Tolos, L., \textit{Hyperonic uncertainties in neutron stars, mergers and supernovae.} Mon. Not.
Roy. Astron. Soc. \textbf{528}, 2629 (2024)

\bibitem{Punturo:2010zz}Punturo, M. et al. \textit{The Einstein Telescope: A third-generation gravitational wave observatory}. Class. Quant. Grav. \textbf{27}  194002 (2010)

\bibitem{LIGOScientific:2016wof}Abbott, B. et al. \textit{Exploring the Sensitivity of Next Generation Gravitational Wave Detectors.} Class. Quant. Grav. \textbf{34}, 044001 (2017)


\bibitem{Blacker:2023opp}Blacker, S., Kochankovski, H., Bauswein, A., Ramos, A. \& Tolos, L. \textit{Thermal behavior as indicator for hyperons in binary neutron star merger remnants.} Phys. Rev. D. \textbf{109}, 043015 (2024)

\bibitem{Alford_2005} Alford, M., Braby, M., Paris, M. \& Reddy, S. \textit{Hybrid Stars that Masquerade as Neutron Stars.} Astrophys. J. \textbf{629}, 969 (2005)


\bibitem{PhysRevD.88.044026} Hotokezaka, K., Kiuchi, K., Kyutoku, K., Muranushi, T., Sekiguchi, Y., Shibata, M. \& Taniguchi, K. \textit{Remnant massive neutron stars of binary neutron star mergers: Evolution process and gravitational waveform}. Phys. Rev. D. \textbf{88}, 044026 (2013) 

\bibitem{refId0}Raduta, A. R., Nacu, F. \& Oertel, M. \textit{Equations of state for hot neutron stars.} Eur. Phys. J. A. \textbf{57}, 329 (2021)


\bibitem{Blacker}Bauswein A. \& Blacker S., \textit{Impact of quark deconfinement in neutron star mergers and hybrid star mergers}.
Eur. Phys. J. \textbf{229}, 3595 (2020) 


\end{thebibliography}
\end{document}